\documentclass[11pt,aps,letterpaper,
preprintnumbers,
amsmath,amssymb,nofootinbib,notitlepage]{revtex4-2}

\usepackage{graphicx}   
\usepackage{longtable}
\usepackage{lineno,hyperref}
\usepackage{amsmath, amssymb}
\usepackage{array,multirow,bigdelim}
\usepackage{xcolor}
\usepackage{easyReview}
\usepackage{ulem}

\begin{document}
	\title{Generalized Langer Correction and the Exactness of WKB \\for all Conventional Potentials}
	
	\author{Asim Gangopadhyaya}
	\email{agangop@luc.edu}
	\affiliation {Department of Physics, Loyola University Chicago, Chicago, IL 60660, U.S.A}
	
	\author{Jonathan Bougie}
	\email{jbougie@luc.edu}
	\affiliation {Department of Physics, Loyola University Chicago, Chicago, IL 60660, U.S.A}
	
	\author{Constantin Rasinariu}
	\email{crasinariu@luc.edu}
	\affiliation {Department of Physics, Loyola University Chicago, Chicago, IL 60660, U.S.A}
	
	\date{\today}
		
\begin{abstract}
	In this paper we investigate the exactness of the WKB quantization condition for translationally shape invariant systems. In particular, using the formalism of supersymmetric quantum mechanics, we generalize the Langer correction and show that it generates the exact quantization condition for all conventional potentials. We also prove that this correction is related to the previously proven exactness of SWKB for these potentials.
	
\end{abstract}

\maketitle

\section{Introduction} WKB is a semiclassical method used to determine approximate eigenvalues and eigenfunctions
\cite{Jeffreys,Wentzel,Kramers,Brillouin,Dunham}.  Even with the theoretical and computational advances of the last hundred years, it is still used in many areas of physics \cite{Weiner1978a, Weiner1978b, Pipard1983, Nieto1985, Poppov1991a, Poppov1991b, Poppov1991c, Dalarsson1993, Maitra1996, Chebotarev1998, Park1998, Price1998, Friedberg2001, Jaffe2010}. Despite the approximate nature of the methodology, surprisingly, this method generates exact spectra for the one-dimensional harmonic oscillator and Morse potentials. However, for other potentials, the WKB method fails to produce the exact eigenvalues or to generate wavefunctions with proper behavior near the singularity. It was soon recognized \cite{Kramers} that with the addition of an \textit{ad hoc} term to the potential, usually known in literature as the Langer correction (although, the more appropriate terminology would be the Kramers or Kramers-Langer correction),  WKB produced exact eigenvalues for the Coulomb and the 3-D oscillator potentials and corrects the behavior of the wavefunction \cite{Langer,Bailey,Froman}.  

Additionally, in the context of supersymmetric quantum mechanics (SUSYQM), it has been shown that a modified version of the method, known as SWKB, generates exact results for an entire class of superpotentials, known as the conventional potentials \cite{Comtet1985,Eckhardt1986,Dutt1986,Raghunthan1987,Dutt1991,Inomata1994,Gangopadhyaya2020}.

 In this paper, we generalize the Langer correction and show that WKB also leads to exact eigenspectra for all conventional potentials with this correction.  This generalized correction and its exactness are closely intertwined with the exactness of SWKB.

The structure of the paper is as follows. In Sec. \ref{back}, we briefly review the formalism of supersymmetric quantum mechanics, the property of shape invariance \cite{Infeld}, and the complete set of conventional potentials. Following previous work \cite{Gangopadhyaya2020}, we group these potentials into three classes based on their functional form. We conclude this section with a brief discussion of WKB method and the Langer correction. In Sec. \ref{glc}, we introduce a generalized form of the Langer correction and show how it applies to each of the three classes of conventional potentials. We then consider each class separately, and prove that with the generalized Langer correction, WKB analysis leads to exact eigenenergies for all conventional potentials. In Sec.~\ref{sec:swkb}, we establish a relationship between our analysis and the previously proven exactness of the SWKB quantization condition for these same potentials. We summarize our work in Sec. \ref{conc}.

\section{Background}\label{back}
\subsection{Supersymmetric Quantum Mechanics}\label{sec:susyqm}
\label{sec:formalism}
Supersymmetric quantum mechanics is a generalization of the ladder operator formulation of the harmonic oscillator \cite{Witten,Solomonson,Dutt_SUSY,CooperFreedman,Cooper-Khare-Sukhatme,Gangopadhyaya-Mallow-Rasinariu}. Analogously, a general hamiltonian $H_- =  -\frac{\hbar^2}{2m} \frac{d^2}{dx^2}+ V_-(x) ~$ is written as a product of two  differential ladder operators
${\mathcal A}^\pm = \mp \, \frac\hbar{\sqrt{2m}}\frac{d}{dx}+ W(x)$ that are hermitian conjugates of each other.   Their product yields
\begin{equation}
	H_-={\mathcal A}^+ {\mathcal A}^-  = \left(  -\hbar \frac{d}{dx}+ W(x) \right)  
	\,\left( \hbar \frac{d}{dx}+ W(x)\right)
	=  -\hbar^2 \frac{d^2}{dx^2}+ V_-(x) ~,
	\label{A+A-}
\end{equation}
where we have set $2m=1$.  The function $W(x)$, known as the superpotential, encapsulates all of the interactions for the system, and it is related to the potential \ $V_-(x)$ by
\begin{equation}
	V_-(x) = W^2(x)  -  \, {\hbar}\, \frac{dW(x)}{dx} ~. \label{eq:Vminus}
\end{equation}
  
By changing the order of the operators ${\mathcal A}^\pm$, we obtain the ``partner'' hamiltonian $H_+= {\mathcal A}^{-}{\mathcal A}^{+}=-\hbar^2 \frac{d^2}{dx^2}+V_+(x)$, where $V_+(x) =
W^2(x) + \hbar \frac{d\, W}{dx}$. These partner hamiltonians are related by 
\begin{equation}\label{eq:intertwining}
 {\mathcal A}^+ H_+ = H_- \, {\mathcal A}^+\quad; \quad {\mathcal A}^- H_-
= H_+\,{\mathcal A}^- ~.
\end{equation}

The hamiltonians $H_\pm$ are semi-positive definite since ${\mathcal A}^\pm$ are hermitian conjugates of each other. Therefore, the eigenenergies $E^{\pm}_{n} \ge 0$. The system is said to have unbroken supersymmetry (SUSY) if either $E^{-}_{0}= 0$ or $E^{+}_{0}=0$. In such cases, we can choose $E^{-}_{0}= 0$ without loss of generality\footnote{If $E^{+}_{0}=0$, we change $W\rightarrow -W$ such that $E^{-}_{0} = 0$; both $E^{-}_{0}$ and $E^{+}_{0}$ cannot be simultaneously zero.}. For systems with unbroken SUSY, Eq.~(\ref{eq:intertwining}) yields 
\begin{eqnarray}
	E^{-}_{n+1} =E^{+}_{n}, \quad ~ n=0,1,2,\cdots~; \label{eq:eigenvalueUB}
\end{eqnarray}
\begin{eqnarray}
	\frac{~~~{\mathcal A}^- }{\sqrt{E^{+}_{n} }} ~\psi^{(-)}_{n+1} 
	= ~\psi^{(+)}_{n} ~~; ~~ ~~
	\frac{~~~{\mathcal A}^+}{\sqrt{E^{+}_{n} }}~\psi^{(+)}_{n} 
	= ~  \psi^{(-)}_{n+1}~. \label{eq:isospectralityUB}
\end{eqnarray}
Thus, except for the ground state, each eigenstate of  $H_-$ shares a common eigenenergy with an eigenstate of $H_+$. For the groundstate with $E^{-}_{0}=0$, ${\mathcal A}^+ {\mathcal A}^- \psi^{(-)}_{0}=0$ leads to ${\mathcal A}^- \psi^{(-)}_{0}=0 $; therefore, $\psi^{(-)}_{0}(x) \sim \exp\left[ -\frac{1}{\hbar}\int^x W(y) \, dy \right]$. 

If neither groundstate energy $E^{\pm}_{0}$ is zero, supersymmetry is broken. In Ref. \cite{Gangopadhyaya2021}, authors showed that unlike in the unbroken case, there are very few systems that hold bound states in the broken supersymmetry phase. In this paper, we will limit our discussion to unbroken supersymmetry.  

If we know the eigenvalues and eigenfunctions of one of the partner hamiltonians, then we can find the same for the other partner, as shown in Eqs. (\ref{eq:eigenvalueUB}) and (\ref{eq:isospectralityUB}). Additionally, if the system possesses a property known as ``shape invariance,'' the ladder operators allow us to generate the entire spectra for both partners, beginning from the  zero-energy groundstate of $H_-$.  

\subsection{Shape Invariance}\label{sec:shapeinvariance}
A superpotential $W(x,a_i)$ is shape invariant
if
\begin{equation}
	W^2(x,a_i)  +  \hbar \frac{d\, W(x,a_i)}{dx}+g(a_i)=
	W^2(x,a_{i+1})  -  \hbar \frac{d\, W(x,a_{i+1})}{dx}+g(a_{i+1})~ ,
	\label{SIC1}
\end{equation}
where $a_{i+1} = f({a_i})$ for functions $f(a)$ and $g(a)$. For the sequence of hamiltonians $H_+\left(a_i\right)$ and $H_-\left(a_{i+1}\right)$, the eigenenergies and corresponding eigenfunctions are related \cite{Infeld,Miller,gendenshtein1,gendenshtein2} as follows:
\begin{eqnarray}
	E_n^{(+)}(a_{i})-E_n^-(a_{i+1})&=&g(a_{i+1})-g(a_{i}),\label{eq:eigenvalues} \\
	\psi^{(+)}_{n}(x,a_{i})&=&
	\psi^{(-)}_n(x,a_{i+1}) \label{eq:eigenfunctions}
	~.
\end{eqnarray}
For a shape invariant superpotential, we can construct the entire spectra of $H_-$ and $H_+$  \cite{Cooper-Khare-Sukhatme,Gangopadhyaya-Mallow-Rasinariu} starting from the groundstate of $H_-\left(a_0\right)$. Using Eqs. (\ref{eq:eigenvalueUB}, \ref{eq:isospectralityUB}, \ref{eq:eigenvalues}), and  (\ref{eq:eigenfunctions}), this procedure yields
\begin{eqnarray}
	E_n^{(-)}(a_0)&=&g(a_n)-g(a_0), \label{eq:En} \\
	\psi^{(-)}_{n}(x,a_0)&=&
	\frac{{\mathcal A}^+{(a_0)} 
		~ {\mathcal A}^+{(a_1)}  \cdots  {\mathcal A}^+{(a_{n-1})}}
	{\sqrt{E_{n}^{(-)}(a_0)\,E_{n-1}^{(-)}(a_1)\cdots E_{1}^{(-)}(a_{n-1})}}~\psi^{(-)}_0(x,a_n)
	~.
\end{eqnarray}
Therefore, shape invariance leads directly to the exact solvability of quantum mechanical systems; i.e., the eigenenergies of the system can be determined as  explicit functions of the parameters of the system and the quantum number $n$. Shape invariance is a symmetry condition intrinsically connected to an underlying potential algebra \cite{Fukui1993,Gangopadhyaya1997,balantekin,Gangopadhyaya1998a, Gangopadhyaya1998b,balantekin2}.

In this paper we consider additive shape invariance described by $a_{i+1}=a_{i}+\hbar$.  
Additive shape invariant superpotentials $W(x,a_i)$ that do not intrinsically depend on $\hbar$ are called  ``conventional".\footnote{Several authors \cite{Ramos00,Ramos99} have called these potentials ``Classical''.}
All conventional superpotentials satisfy the following set of partial differential equations \cite{Gangopadhyaya2008,Bougie2010,symmetry}
\begin{eqnarray}
	W \, \frac{\partial W}{\partial a} - \frac{\partial W}{\partial x} + \frac12 \, \frac{d g(a)}{d a}= 0~, \label{PDE1} \\
	\frac{\partial^{3}}{\partial a^{2}\partial x} ~W(x,a)= 0~.\label{PDE2}
\end{eqnarray}
It follows that the list of superpotentials in  Refs. \cite{Infeld,Dutt_SUSY} is complete; i.e., no other conventional superpotentials exist \cite{Bougie2010,symmetry}.
Additional ``extended'' shape invariant superpotentials have been found \cite{Quesne1,Quesne2,Quesne2012a,Quesne2012b,Odake1,Odake2,Tanaka,Odake3,Odake4}, that each consist of a conventional superpotential plus an $\hbar$-dependent extension \cite{Bougie2010,symmetry}.

\subsection{The complete list of conventional potentials}\label{sec:completelist}
\label{Sec:CompleteList}
In this section,  we list the complete set of conventional potentials. 
As initially conjectured in  Ref. \cite{Infeld} and proven in Ref. \cite{Bougie2010},  all conventional superpotentials must be of the form 
\begin{equation}W(x,a) = a f_1(x) + f_2(x) +u(a)\label{Wform}\end{equation}
As a consequence, conventional potentials break into  three classes: Class I ($f_1=$ constant), Class II ($f_2=$ constant), and Class III (neither $f_1$ nor $f_2$ is constant) \cite{symmetry}.

\subsubsection{Class I Potentials}
\label{Sec:classI}
For this class  $f_1=\alpha$, for some constant $\alpha$. In this case, $W(x,a)=f_2(x)+\alpha\,a$, where $f_2' = \alpha f_2 - \varepsilon$. This results in two subclasses: Class IA, where $\alpha = 0$, and Class IB, where  $\alpha\neq0$. These superpotentials and their corresponding potentials are given in Table \ref{table:conventional-I}.

\begin{table} [htb]
	\begin{center}
		\begin{ruledtabular}
			\begin{tabular}{llll}
				\textbf{Class}& \textbf{Superpotential $W$}&\textbf{Potentials $V_{\pm}$ }&\qquad \textbf{Name}\\ 
				\hline
				IA &$\frac12 \omega x$ & $\frac14 \,\omega ^2x^2 \pm \frac12 \, \hbar \omega$ &Harmonic Oscillator\footnote{For the harmonic oscillator potential, $W$ and $dg/da$ are independent of $a$.}  \\
				($\alpha=0$)& $\left( \varepsilon = -\frac 12 \omega\right)$&& \\
				\hline
				IB &$A-e^{-x}$&$A^2-\left( 2 A \mp \hbar \right) e^{-x}+e^{-2 x}$ &Morse\footnote{For the Morse potential, $\varepsilon=0$ and $\alpha $ is a scaling factor; we choose $\alpha = -1$ to agree with existing literature.}\\
				($\alpha=-1$)&$\left(a=-A\right)$&& \\
			\end{tabular}
		\end{ruledtabular}
		\caption{Class I superpotentials are of the form $W(x,a)=f_2(x)+\alpha\,a$, falling into two subclasses according to the value of $\alpha$. For each subclass, this table lists $W$ with the appropriate values of parameters $a$ and $\varepsilon$, and the corresponding potentials $V_{\pm}$.}
\label{table:conventional-I}
	\end{center}		
\end{table}

\subsubsection{Class II Potentials}
\label{Sec:classII}
For Class II, $W(x,a)=a f_1(x)+B/a$, $f_1'=f_1^2-\lambda$, and $W^\prime(x,a)=a f_1'(x)$; this produces two subclasses: Class IIA, where $\lambda=0$, and Class IIB, where $\lambda\neq 0$. Class IIB further divides based on the signs of $\lambda$ and $a$. These superpotentials and their corresponding potentials are given in Table \ref{table:conventional-II}.

\begin{table} [htb]
	\begin{center}
		\begin{ruledtabular}
			\begin{tabular}{llll}
				\textbf{Class}& \textbf{Superpotential $W$}&\textbf{Potentials $V_{\pm}$ }&\qquad \textbf{Name}\\ \hline
				IIA&$-\frac \ell r+ \frac{e^2}{2\ell}~$ & $-\frac{e^2}{r}+\frac{\ell(\ell \pm \hbar)}{r^2} +\frac{e^4}{4 l^2}$ & Coloumb\\ 
				($\lambda=0$)&$\left( B = \frac{1}{2}e^2\, , ~a=\ell, \quad f_1=-\frac1r\right)$&& \\
				\hline
				IIB1&$-A\,\cot x - \frac BA$& $A \left(A\pm\hbar \right)  \csc ^2x +$ & Rosen-Morse  \\
				($\lambda<0$)& $\left(a=A, \quad f_1=-\cot x\right)$ & $2 B \cot x+\frac{B^2}{A^2}- A^2$& (Trigonometric)\\
				\hline
				IIB2&$ A\,\tanh x + \frac BA~$&
				$-A \left(A\mp\hbar \right)  \text{sech}^2x+$&Rosen-Morse \\
				($\lambda>0$, $a<0$) &$\left(a=-A,~B\to -B, \quad f_1=-\tanh x\right)$ &$2 B \tanh x+\frac{B^2}{A^2}+A^2$& (Hyperbolic) \\
				\hline
				IIB3&$-A\coth r + \frac BA~$& $A (A\pm \hbar)\, \text{csch}^2r-$   &Eckart\\
				($\lambda>0$, $a>0$) &$\left(a=A, \quad f_1=-\coth r\right)$&$2 B \coth r  + \frac{B^2}{A^2}$& \\
			\end{tabular}
		\end{ruledtabular}
		\caption{Class II superpotentials are of the form $W(x,a)=af_1(x)+B/a$, falling into subclasses according to the values of $\lambda$ and $a$. For each subclass, this table lists $W$ and the corresponding potentials $V_{\pm}$.}	
		\label{table:conventional-II}
	\end{center}
\end{table}

\subsubsection{Class III Potentials}
\label{Sec:classIII}

For Class III, $W(x,a)=a f_1(x) + f_2(x)$. Here $f_1^\prime=f_1^2 -\lambda$ and $f_2^\prime = f_1 f_2- \varepsilon$, where $\lambda$ and $\varepsilon$ are constants. 
This class further splits into two subclasses:  Class IIIA, where $\lambda= 0$,  and Class IIIB (which further divides three ways), where $\lambda\neq 0$. 
These superpotentials and their corresponding potentials are given in Table \ref{table:conventional-III}.
\begin{table} [h!]
	\begin{center}
		\begin{ruledtabular}
			\begin{tabular}{llll}
				\textbf{Class}& \textbf{Superpotential $W$}&\textbf{Potentials $V_{\pm}$ }&\qquad \textbf{Name}\\ \hline
				IIIA&$\frac12\, \omega r -\frac \ell r$& $\frac14\, \omega ^2\,r^2 +\frac{\ell(\ell\pm\hbar)}{r^2}-\left( \ell \mp  \frac{h }{2} \right) \omega$  & 3D-Oscillator\\
				($\lambda=0$)&$\left( a= \ell, \quad f_1=-\frac1r\right)$&&\\
				\hline
				IIIB1&$A\tan x - B\, {\rm sec\,}x$ & $\left(A (A\pm \hbar)+B^2\right)\sec^2x -$ & Scarf  \\	
				($\lambda<0$)&$\left(a=A, \quad f_1=\tan x\right)$&$B (2 A\pm\hbar) \tan x \sec x -A^2$& (Trigonometric)\\
				\hline
				IIIB2&$A\tanh x + B \, {\rm sech\,}x$  &
				$- \left(A(A \mp h)-B^2\right){\rm sech}^2x+$&Scarf \\
				($\lambda>0$, $f_1^2<\lambda$) &$\left(a=-A, \quad f_1=-\tanh x\right)$&$B (2 A\mp h) \tanh x \,{\rm sech}\,x+A^2$& (Hyperbolic)\\
				\hline
				IIIB3& $A\coth r - B \, {\rm csch\,}r$ & $ \left(A(A\mp  \hbar)+B^2\right)\text{csch}^2r-$ &P\"oschl-Teller \\
				($\lambda>0$, $f_1^2>\lambda$) &$\left(a=-A, \quad f_1=-\coth r\right)$&$B (2 A\pm \hbar) \coth r \,\text{csch}\,r+A^2$& (Hyperbolic) \\
			\end{tabular}
		\end{ruledtabular}
		\caption{Class III superpotentials are of the form $W(x,a)=af_1(x)+f_2(x)$, falling into subclasses according to the value of $\lambda$ and the sign of $f_1^2-\lambda$. This table lists $W$ and the corresponding potentials $V_{\pm}$.}
		\label{table:conventional-III}
	\end{center}
\end{table}

\subsection{WKB Quantization and Langer Correction}
\label{sec:Exact}
The WKB formalism is a simple, physically intuitive, non-perturbative approximation method for finding eigenvalues, eigenfunctions, and transitional probabilities for quantum mechanical systems.  In this formalism, the eigenvalues are determined from the quantization condition
\begin{equation}
	\int_{x_L}^{x_R} \sqrt{E_n-V(x)} ~dx = \left( n+\nu \right) \pi \hbar~, \label{eq.Bohr-Sommerfeld}
\end{equation}
where $\nu$ is a fractional number and depends on the nature of the potential. The limits of integration ${x_L}$ and ${x_R}$ are the classical turning points; i.e., solutions of ${E_n-V(x)}=0$. The corresponding eigenfunctions are built by patching together functions defined within classically allowed and forbidden regions using connection formulas.

The quantization condition of Eq. (\ref{eq.Bohr-Sommerfeld}) is supposed to give excellent approximations for large values of $n$, but for certain smooth potentials, the validity of this method can also be extended \cite{Karnakov} to $n\sim 1$.  Surprisingly, it has been shown to produce exact eigenvalues for both one-dimensional harmonic oscillator and Morse potentials with $\nu = \frac12$. However,  aside from the Class I potentials, the WKB method does not give exact results for other conventional potentials.

Very early on, Kramers \cite{Kramers} recognized that for the Coulomb potential, WKB failed to capture the proper behavior of the radial part of the wave function  near the origin. Kramers also noted that if the centrifugal term in the potential, $\frac{\ell(\ell+\hbar)}{r^2}$, were to be modified to $\frac{(\ell+\frac\hbar2)^2}{r^2}$, the resulting wave functions picked up the correct $r^{\ell+1}$ behavior as $r \rightarrow 0$. This is equivalent to adding a term $\frac14 \frac{\hbar^2}{r^2}$  to the potential. Young and Uhlenbeck \cite{Young1930} demonstrated that this replacement also gives the correct eigenenergies.

Langer later \cite{Langer1949} extended this analysis to the radial oscillator, and the additional term became known as the \textit{Langer correction}. Note that the Langer correction accomplishes two tasks in the cases of the radial oscillator and Coulomb potentials. First,  it matches the behavior of the wavefunction in the asymptotic limits. Second, it produces the exact energy eigenvalues using the Bohr-Summerfeld quantization condition.

Langer \cite{Langer} justified  the additional term by change of variable $r= e^x$ and change of wave function $\psi=e^{x/2}u$. 
Langer also argued that the inexactness of WKB for the Coulomb  potential emanated from the semi-infinite nature of its domain, and WKB analysis was suited for potentials defined over the entire real axis. Here, the changing of domain from semi-infinite to infinite resulted in the change from $\ell(\ell+\hbar)$ to $(\ell+\hbar/2)^2$ in these two potentials. However, it has been noted \cite{Engelke1970,Adams1977,Dahl2004,Koike2009,Li2020} that this change of variables is not unique, and 
some have argued \cite{Froman2004} that the mapping from semi-infinite to infinite domain is not fundamental to the success of the Langer correction in accomplishing the two tasks listed above.
	In this paper we introduce a generalized Langer correction. This correction is introduced heuristically in the spirit of Kramers to accomplish the two tasks listed above, rather than from a Langer-like coordinate transformation.

	In a different approach to semiclassical methods, it has been observed \cite{Comtet1985,Eckhardt1986,Dutt1986,Dutt1991,Inomata1994,Gangopadhyaya2020} that a modified version of WKB known as SWKB applies to the superpotential $W(x,a)$ in the context of SUSYQM.
	This condition states that 
	\begin{equation}\label{eq:swkb}
	\int_{x_L}^{x_R} \sqrt{E_n(a)-W^2(x,a)} ~dx = n \hbar \pi~.
	\end{equation}
	It has previously been shown on a case-by-case basis that this SWKB condition is exact for all conventional superpotentials and leads to the correct asymptotic behavior of the wavefunction, without the need for a Langer-type correction \cite{Adhikari1988}. The authors of \cite{Gangopadhyaya2020} proved that this exactness follows directly from the form of the potentials required by shape-invariance. 
	
	Returning to the WKB formalism, several authors \cite{Rosenzweig1968,Li2020} have introduced different versions of the Langer correction, and previous authors \cite{Sun2012} have shown on a case-by-case basis that changes in parameters can lead to exactness for all conventional potentials.
	In the present work, we show that a generalized correction $\frac{\hbar^2}{4}f_1^\prime$ provides a universal recipe that ensures that WKB generates exact eigenvalues for all conventional potentials and regularizes the wavefunction near any singularities.

\section{Generalized Langer Correction} \label{glc}

As shown in Eq.~\ref{Wform}, all conventional potentials are of the form 
$W(x,a) = a\, f_1(x)+f_2(x)+u(a)~,$
and they break down into three classes. The quantization condition of Eq. (\ref{eq.Bohr-Sommerfeld}) is exact for Class I potentials, where $f_1$ is constant, without the need for a Langer correction. On the other hand, the Coulomb and 3D-Oscillator potentials both have $f_1 = -\frac1r$, and the corresponding required Langer corrections are $\frac{\hbar^2}4\, \frac1{r^2}=\frac{\hbar^2}{4}f_1^\prime$, as shown in Table \ref{table:Langer}.

\begin{table*} [h!]
	\begin{center}
		\begin{ruledtabular}
			\begin{tabular}{lllrll}
				Potential & $W$ & $f_1$  && \hspace{-0.2cm} Langer correction& Generalized form \\ 
				\hline
				{Coulomb}& {$a f_1(x)+B/a$} & {$f_1 = - \frac1r $}& & $\frac{\hbar^2}4\, \frac1{r^2}$& $\frac{\hbar^2}{4}f_1^\prime = \frac{\hbar^2}4\, \frac1{r^2}$ \\
				{Radial Oscillator}& {$a f_1(x)+f_2(x)$} &{$f_1 = - \frac1r $}&
				&$\frac{\hbar^2}4\, \frac1{r^2}$&$\frac{\hbar^2}{4}f_1^\prime = \frac{\hbar^2}4\, \frac1{r^2}$\\
			\end{tabular}
		\end{ruledtabular}
		\caption{Standard Langer Correction for the radial potentials.}
		\label{table:Langer}
	\end{center}
\end{table*}

Our search for a generalized Langer correction was guided by the following three heuristics. 
\begin{itemize}
\item The correction had to be universal; i.e., when applied to all conventional potentials, it had to produce the correct energy eigenvalues. 
\item It had to reproduce the known Langer corrections for the Coulomb and the radial oscillator potentials, while producing no correction for the simple harmonic oscillator and Morse. 
\item It had to regularize the wavefunction near any singularity of the potential; i.e., the wavefunction must have a correct asymptotic form as required by the Schr\"odinger equation. 
\end{itemize}
In this section, we report on a generalized Langer correction and prove that it fits the above heuristics. 

The conventional superpotentials are of the form $W = a f_1 + f_2 + u$. For the Coulomb and 3D-oscillator potentials we observe that the Langer correction can be written as $\frac{\hbar^2}{4}f_1^\prime$. Additionally, for Class I, where no correction is needed, $f_1'=0$. Based on these observations, we hypothesize a generalized Langer correction of the form $\frac{\hbar^2}{4}f_1^\prime$. Consequently, the generalized Langer-corrected potential becomes
\begin{equation}
\label{eq:gen-Langer}	
V(x)=V_-(x) + \frac14 \hbar^2 f_1^\prime(x)~.
\end{equation}
In Table \ref{table:classes}, we list these generalized corrections for all three classes.
\begin{table} [htb]
	\begin{center}
		\begin{ruledtabular}
			\begin{tabular}{lllrll}
				Class&Form of $W$ & Constraints from && \hspace{-0.2cm}Subclasses & Generalized Langer    \\ 
				&&shape invariance &&&Correction $\frac{\hbar^2}{4}f_1^\prime $\\ \hline
				\multirow{2}{*}{Class I}& \multirow{2}{*}{$f_2(x)+\alpha\, a$} & \multirow{2}{*}{$\alpha f_2-f_2'=\varepsilon$ }& \multirow{2}{*}{$\Biggl\lbrace$}&\hspace{-0.15cm}IA: $\alpha=0$&$\frac{\hbar^2}{4}f_1^\prime = 0$\\
				&&&&\hspace{-0.15cm}IB: $\alpha\neq0\hspace{1cm}$  &$\frac{\hbar^2}{4}f_1^\prime = 0$\vspace{0.5cm}\\
				\multirow{2}{*}{Class II}& \multirow{2}{*}{$a f_1(x)+B/a$} & \multirow{2}{*}{$f_1^2-f_1'=\lambda$}&
				\multirow{2}{*}{$\Biggl\lbrace$}&\hspace{-0.15cm}IIA: $\lambda=0$& $\frac{\hbar^2}{4}f_1^\prime = \frac{\hbar^2}{4} \, f_1^2$\\
				&&&&\hspace{-0.15cm}IIB: $\lambda\neq0$&$\frac{\hbar^2}{4}f_1^\prime = \frac{\hbar^2}{4} \, \left( f_1^2 -\lambda\right) $\vspace{0.5cm}\\
					\multirow{2}{*}{Class III}& \multirow{2}{*}{$a f_1(x)+f_2(x)$} &$f_1^2-f_1'=\lambda$,&
				\multirow{2}{*}{$\Biggl\lbrace$} &\hspace{-0.15cm}IIIA:$\lambda=0$ &$\frac{\hbar^2}{4}f_1^\prime = \frac{\hbar^2}{4} \, f_1^2$\\
				&&$f_1f_2-f_2'=\varepsilon$&&\hspace{-0.15cm}IIIB:$\lambda\neq0$&$\frac{\hbar^2}{4}f_1^\prime = \frac{\hbar^2}{4} \, \left( f_1^2 -\lambda\right) $\\
			\end{tabular}
		\end{ruledtabular}
		\caption{Three classes of conventional shape invariant superpotentials and their properties. The following are all constants: $\alpha$, $\varepsilon$, $\lambda$, and $B$.}
		\label{table:classes}
	\end{center}
\end{table}

We first show that this generalized correction regularizes the wavefunctions near the singularity. We then prove that this addition assures that every conventional potential satisfies the  Bohr-Sommerfeld quantization condition 
\begin{equation}\label{eq:wkb-1}
	\int_{x_L}^{x_R} \sqrt{E_n(a)-\left(V_-(x,a) + \frac14 \hbar^2 f_1'(x)\right) 
	}\quad {\rm d}x  = \left(n+\frac12 \right)\pi \hbar ~,
\end{equation}
where $x_L$ and $x_R$ are the turning points for the modified potential. 
This is the quantization condition of Eq. (\ref{eq.Bohr-Sommerfeld}), with $\nu=1/2$,  using the corrected  potential from Eq. (\ref{eq:gen-Langer}).

\bigskip
\noindent
\textbf{Wavefunctions Near Singularity}

Here we prove that our generalized correction regularizes the wavefunctions near the singularity. 
Class I potentials do not have a singularity within the domain, and do not need any correction; this is consistent with the condition $f_1'=0$ for this class, as shown in Table \ref{table:classes}. On the other hand, $f_1$ is not constant for Classes II and III.

For Class II superpotentials with singularity, the superpotentials are of the form $W= af_1+B/a$, where $a$ is positive. In the vicinity of a singularity, $|f_1|\rightarrow \infty $ and $f_1^\prime = f_1^2-\lambda \approx f_1^2$. Then, the  Schr\"odinger equation reads
\[
-{\hbar^2}
\frac{d^2 \psi}{dx^2} 
+a\left( a -  \hbar \right) f_1^2 \,\psi + 2B f_1\, \psi +\frac{B^2}{a^2} \,\psi= E\, \psi
~.
\]
Since the Bohr-Sommerfeld condition requires two turning points, and since $f_1$ is monotonic, we must have $a(a-\hbar)>0$ and $-{\hbar^2}
	\frac{d^2 \psi}{dx^2} 
	+a\left( a -  \hbar \right) f_1^2 \,\psi  \approx 0$.
Since $E \ll f_1^2$ near the singularity, the wave function in this region is given by the zero-energy solution 
	$$\psi \sim e^{-\frac1\hbar \int^x W(x,a)dx} 
	\approx e^{-\frac1\hbar \int^{f_1} \left(af_1+B/a\right) df_1/f_1^2 } 
	\approx e^{-\frac{a}\hbar \int^{f_1}  df_1/f_1 }~
	.$$
Hence, $\psi \sim f_1^{-a/\hbar}$.  As an illustration, for the Coulomb potential, $f_1=-1/r$ and $a/\hbar=\ell+1$; the wave function becomes $\psi \sim r^{\ell+1}$.

In WKB, the wave function near singularity is given by 
$$
\psi =\frac1{\sqrt{Q}} e^{-\frac1\hbar \int^x Q\,dx }~,
$$
where $Q=\, \sqrt{V-E}$. For the uncorrected potential,
$$
Q=\, \sqrt{V_- -E}=\, \sqrt{W^2-\hbar W^\prime-E}\,\approx\,  |f_1|\,\sqrt{a(a-\hbar)}~ ~,
$$
which requires $a>\hbar$. The resulting wavefunction does not have the correct behavior; i.e., $\psi \sim f_1^{-a/\hbar}$, which is fixed by the generalized Langer correction. Adding $\frac14 \hbar^2 f_1^\prime$ to the potential gives
\begin{equation}
	\label{eq:a>h}
	\tilde{Q}\approx (a-\hbar/2)\,|f_1|~.
~
\end{equation}
Then
\[
\int^x \tilde{Q}\, dx = (a-\hbar/2) \int^x |f_1| \, dx = (a-\hbar/2) \int^{|f_1|} |f_1| \, \frac{d|f_1|}{f_1^2} = (a-\hbar/2) \ln |f_1|~,\]
which yields
\[
\tilde{\psi} \rightarrow \frac{1}{\sqrt{|f_1|}} e^{-(a-\hbar/2) \ln |f_1|} = |f_1|^{-\frac a\hbar +\frac12 -\frac12}
= |f_1|^{-\frac a\hbar}~.
\]
Returning to the example of the Coulomb potential  we have $\tilde{\psi}\to r^{\frac a\hbar} =r^{\ell+1}$.

For Class IIIA, $W \to a f_1$ near singularity, hence the analysis of Class II holds.  
For Class IIIB, near a singularity, 	
	$W = af_1+f_2 = af_1+B\sqrt{|f_1^2-\lambda|} \approx (-a+B)|f_1|$.  For $a>0$, we obtain  
\begin{equation}
	\label{eq:a>hB}
	\tilde{Q}\approx (a-|B|-\hbar/2)\,|f_1|~,
\end{equation}
provided that $(a-|B|-\hbar/2)>0$. For $a<0$ and $B<0$ 
\begin{equation}
	\label{eq:a>hC}
	\tilde{Q}\approx (a-B-\hbar/2)\,|f_1|~
\end{equation}
provided that $(a-B-\hbar/2)>0$.

\bigskip
\noindent
\textbf{Quantization Conditions}\label{subsec:Calc}

In the following we will prove that the conventional potentials with the generalized Langer correction produce the exact eigenspectra for these systems. We define an integral $ I(a,n,\hbar)$:
\begin{equation}
 I(a,n,\hbar) \equiv \int_{x_1}^{x_2} \sqrt{E_n-\left( W^2(x,a) - \hbar W^\prime + \frac14 \hbar^2 f_1^\prime\right) 
	}\quad {\rm d}x ~,\label{eq:wkb0}
\end{equation}
and show that 
\[
I(a,n,\hbar)  = \left( n+ \frac{1}{2} \right) \hbar \pi~
\] 
by separately considering each of the three classes.

\subsection{Class I}\label{subsec:classI}
For a superpotential of Class I, $f_1(x) = \mu$, a constant, and therefore, after a proper reparametrization \cite{Gangopadhyaya2020} we can write  $W(x,a) = f_2(x) + \alpha\, a$ where $f_2'(x) = \alpha f_2(x) -\varepsilon$. Therefore, the generalized Langer correction, $\frac{\hbar^2}{4}f_1^\prime$, is zero for this class.

The potential $V_- = W^2- \hbar W'$ is
\begin{equation}
	\label{eq:V-1}
	V_-(x,a)=  \alpha ^2 a^2 + \alpha (2 a-\hbar) f_2(x)+f_2(x)^2+\epsilon \hbar~.
\end{equation}
From Eq. \ref{PDE1}, we get
\begin{equation}
	\label{eq:classI-dgda}
	\frac{dg(a)}{da}=-2\, ( \alpha ^2 a+\varepsilon )\quad \text{i.e.,}\quad
	g(a) = -a \,( \alpha ^2 a +2 \varepsilon) ~,
\end{equation}
and consequently, the eigenenergies are 
\begin{equation}
	\label{eq:En-SI-caseI}
	E_n = g(a+n\hbar) - g(a) = -n \hbar \left(2 \alpha ^2  a +\alpha ^2 n \hbar +2 \varepsilon \right)~.
\end{equation}
To avoid energy level crossing, we must have $dg/da > 0$. 

This class splits into Subclass IA, with $\alpha = 0$ and Subclass IB, with $\alpha \ne 0$. For $\alpha=0$, $f_2' = -\varepsilon$ and for $\alpha \ne 0, \varepsilon = 0$, so \cite{Gangopadhyaya2020} $f_2' = \alpha f_2$. In either case, $W' = f_2'$ cannot cross zero; consequently $f_2'$ has a definite sign, which must be positive for unbroken supersymmetry. Thus, $f_2$ is a monotonically increasing function.

\subsubsection{Subclass IA: $\alpha=0$}\label{sec:IA}
We consider first Subclass IA, for which $\alpha=0$. Therefore, $W = f_2$, $f_2' =-\varepsilon$,  $E_n = -2n \hbar \varepsilon$, and the integral (\ref{eq:wkb0}) is
\begin{equation}
	\label{eq:wkb-I}
	I(a,n,\hbar) = \int_{x_1}^{x_2} 
	\sqrt{-f_2^2(x)- (2 n  +1 )\hbar\,\varepsilon}\, dx~.
\end{equation}
Because $f_2$ is monotonic, we change the integration variable to $f_2$, and  obtain
\begin{equation}
	\label{eq:wkb-Ia}
	I(a,n,\hbar) =  \int_{f_{2L}}^{f_{2R}} 
	\sqrt{-f_2^2(x)- (2 n  +1 )\hbar\,\varepsilon}~ \frac{ df_2}{-\varepsilon}~.
\end{equation}
The integral limits $f_{2L} = -\sqrt{-2 n\hbar \epsilon -\hbar \epsilon }$ and $f_{2R}=\sqrt{-2 n\hbar \epsilon -\hbar \epsilon }$ correspond to the classical turning points of the particle with energy $E_n$. 

We can write this in the form of integral $I_0(y_1,y_2)$ in the Appendix,
\begin{equation}
	\label{eq:1a}
	I(a,n,\hbar)=\int_{f_{2L}}^{f_{2R}}
	\sqrt{(f_{2R}-f_2)(f_2-f_{2L})}\,\,\frac{ df_2}{-\varepsilon}~,
\end{equation}
which yields
\begin{equation}
	\label{eq:wkb-Iaa}
	I(a,n,\hbar) = \left(n+\frac 12 \right)\hbar \pi~.
\end{equation}

\subsubsection{Subclass IB: $\alpha \ne 0$}\label{sec:IB}

In this case $W = f_2 + \alpha a$ and $\alpha f_2 - f_2' = \varepsilon$. Note that we can set $\varepsilon = 0$ by redefining the shape invariance parameter $a + \varepsilon/a \to a$.  Then, $E_n = -\alpha ^2  n \hbar\, (2 a+n\hbar)$, and therefore, the WKB integral (\ref{eq:wkb0})  becomes
\begin{equation}
	\label{eq:wkb-IB}
	I(a,n,\hbar) = \int_{x_1}^{x_2} 
	\sqrt{-a^2\alpha^2 +(\hbar-2 a) f_2(x)-\alpha ^2 n \hbar \, (2 a+ n\hbar)-f_2(x)^2}\, dx~.
\end{equation}
From Table 	\ref{table:classes}, the generalized Langer correction is zero. The limits $x_1$ and $x_2$ are the turning points for the classical particle with energy $E_n$. Due to the monotonic behavior of $f_2$ we change the integration variable $dx = df_2/f_2' =  df_2/(\alpha f_2)$. Then, Eq. (\ref{eq:wkb-IB}) becomes
\begin{equation}
	\label{eq:wkb-IB2}
	I(a,n,\hbar) = \int_{f_{2L}}^{f_{2R}} 
	\sqrt{-a^2 \alpha^2+ (\hbar-2 a) f_2 -\alpha ^2 n\hbar (2 a+n\hbar)-f_2^2}~ \frac{df_2}{\alpha  f_2}\, ~.
\end{equation}
In the new variable $f_2$, the turning points $f_{2L}$ and $f_{2R}$ are the roots of 
\begin{equation}
	\label{eq:wkb-IB5}
	-a^2+(\hbar-2 a) f_2(x)-\alpha ^2 n \hbar \, (2 a+ n\hbar)-f_2(x)^2 = 0~,
\end{equation}
i.e.;
\begin{equation}
	\label{eq:wkb-IB6}
	\left( \begin{array}{c} f_{2L} \\ f_{2R} \end{array}\right) 
	= 	\frac{1}{2} \left(\alpha \hbar -2 a \alpha \mp \sqrt{\alpha^2 \hbar ^2 - 8  \alpha^2 a n \hbar - 4 \alpha^2 a n \hbar -4 \alpha^2 \hbar^2 n^2} \right)~.
\end{equation}
The integral of Eq. (\ref{eq:wkb-IB2}) can be written as
\begin{equation}
	\label{eq:wkb-IB7}
	I(a,n,\hbar) =\frac{1}{\alpha}\int_{f_{2L}}^{f_{2R}}
	\sqrt{(f_{2R}-f_2)(f_2-f_{2L})}\,\, \frac{df_2}{f_2}~.
\end{equation}
Because $f_{2L}<f_{2R}<0$, this integral is of the form of $I_{1b}(y_1,y_2)$ in the Appendix. Therefore 
\begin{equation}
	\label{eq:wkb-IB8}
	I(a,n,\hbar) =\frac{\pi}{2\alpha }\left( f_{2L}+f_{2R}\right) + \frac{\pi}{\alpha}\sqrt{f_{2L} \, f_{2R}}~.
\end{equation}
Using Eq. (\ref{eq:wkb-IB6}), we obtain after simplifications:
\begin{equation}
	\label{eq:wkb-IBfinal}
	I(a,n,\hbar) = \left(n+\frac 12 \right)\hbar \pi~.
\end{equation}

\subsection{Class II}\label{subsec:classII}

For this class the superpotential is of the form $W(x,a)=a f_1(x)+B/a$, where $B$ is a constant.
From Eq. \ref{PDE1}, we have
\begin{equation}
	\label{eq:f2const}
	\frac{dg}{da} = \frac{2B^2}{a^3} - 2 \lambda \, a; \quad\mbox{i.e.,}\quad  g(a)=-\frac{B^2}{a^2} - \lambda\, a^2 ~,
\end{equation}
and hence the energy eigenvalues are
\[
E_n=\frac{B^2}{a^2} - \frac{B^2}{(a+n\hbar)^2} +\lambda\left[\, a^2-(a+n\hbar)^2\right]
~.
\]
In order to avoid level-crossing, we must have $dg/da = \frac{2B^2}{a^3} - 2 \lambda \, a>0$.  Therefore, if $\lambda\leq 0$, we must have $a>0$. For $\lambda>0$, there are two possibilities:  $a>0$ and ${B^2} >  \lambda \, a^4$, or  $a<0$ and ${B^2} <  \lambda \, a^4$.  Furthermore, since $W' = a f_1' = a (f_1^2-\lambda)$ and $f_1^2$ can never equal\footnote{Otherwise, $f_1^2$ must equal $\lambda$ for all values of $x$.} $\lambda$, $W'$ must have a fixed sign. For unbroken supersymmetry, this sign must be positive. Hence, $a>0$ corresponds to cases in which $f_1^2>\lambda$, and  $a<0$ corresponds to $f_1^2 <\lambda$.

\subsubsection{Subclass IIA: $\lambda=0$}\label{sec:IIA}
In this case $f_1' = f_1^2$. Therefore, if $f_1=0$ at one point, it must be zero at all points.  Hence $f_1$ must have a definite sign everywhere. We choose $f_1<0$ without loss of generality. Then, since $W = af_1 + B/a$ must change sign to preserve supersymmetry, and since $a>0$, we must have $B>0$. Changing the integration variable from $x$ to $f_1$, and using 
$f_1^\prime = f_1^2$ together with $W' =  a \,f_1^2 $, Eq. (\ref{eq:wkb0}) becomes
\begin{equation}\label{eq:I_classIIA}
	I(a,n,\hbar)= 
	\int_{f_{1L}}^{{f_{1R}}}
	\sqrt{-\left( a - \frac h2 \right)^2  f_1^2-2 B f_1-\frac{B^2}{(a+h n)^2}}\quad 
	\frac{df_1}{f_1^2}~~.
\end{equation}
The limits of the integral are the solutions of
\[
-\left( a - \frac h2 \right)^2  f_1^2-2 B f_1-\frac{B^2}{(a+h n)^2} = 0~,
\]
i.e;
\begin{equation}\label{eq:TurningPts_classIIA}
	\left( \begin{array}{c} f_{1L} \\ f_{1R} \end{array}\right) 
	= \frac{2 \left(\mp \frac{B \sqrt{h} \sqrt{2 n+1} \sqrt{4 a+2 h n-h}}{a+h n}-2 B\right)}{\left( 2a-\hbar \right)^2} ~.
\end{equation}
As shown in Eq. (\ref{eq:a>h}), this case requires $a > \hbar/2$ for the proper behavior of the wave function near the singular points. 

Since the integrand of Eq. (\ref{eq:I_classIIA}) is real and positive between the turning points, we have
\begin{equation}\label{eq:I_classIIA2}
	I(a,n,\hbar)= \left( a - \frac h2 \right)
	\int_{f_{1L}}^{{f_{1R}}} ~
	\sqrt{\left(    f_{1R} - f_1 \right) \left(    f_1 - f_{1L}  \right)   }
	\quad 
	\frac{df_1}{f_1^2}~~.
\end{equation}  
This integral is of the form of $I_{2a}(y_1,y_2)$ in the Appendix. Thus we obtain
\begin{equation} 
	I(a,n,\hbar)= -\pi  \,\left( a - \frac h2 \right) \frac{{f_{1R}}\,  \sqrt{{f_{1L}} \, {f_{1R}} }+{f_{1L}}  \left(\sqrt{{f_{1L}} \, {f_{1R}} }+2\,  {f_{1R}} \right)}{2\,  {f_{1L}} \, {f_{1R}} }~,
\end{equation}
which, after substitution of the turning points from Eq. (\ref{eq:TurningPts_classIIA}), yields 
\begin{equation} 
	I(a,n,\hbar)= \left( n+ \frac{1}{2} \right) \hbar \pi~.
\end{equation}

\subsubsection{Subclass IIB: $\lambda\neq0$}\label{sec:IIB}
The superpotentials for this subclass are of the form
\[W(x,a) = a f_1(x) +B/a~,
\]
where $f_1$ obeys the equation
$
f_1^2-f_1^\prime =\lambda ~. 
$
For $\lambda \neq 0$, we can choose $|\lambda|=1$ by scaling $x$ and $f_1$. Then, the potential $	V_- = W^2-\hbar W^\prime$  becomes
\begin{equation}
	V_-(x) = a^2 f_1(x)^2-a \hbar  f_1(x)^2+a \lambda  \hbar +2 B f_1(x)+\frac{B^2}{a^2}~,
\end{equation}
and the corresponding Langer correction is $\frac14 \hbar^2 f_1^\prime = \frac14 \hbar^2 \left( f_1^2-\lambda\right)$. The modified potential including the Langer term is then given by
\begin{equation}
	\label{eq:L}
	V(x) = \left( a-\frac\hbar 2\right) ^2 f_1(x)^2+a \lambda  \hbar +2 B f_1(x)-\frac{\lambda  \hbar ^2}{4}+\frac{B^2}{a^2}~.
\end{equation}
Hence,  after 
substituting $W' = a f_1' = a(f_1^2-\lambda)$ and changing variable from $x$ to $f_1$, Eq. (\ref{eq:wkb0}) becomes
\begin{equation}\label{eq:I_classIIB}
	I(a,n,\hbar)= 
	\int_{f_{1L}}^{{f_{1R}}}\,{df_1}\, 
	\frac{\sqrt{
			-\left( a-\frac\hbar2\right)^2 f_1^2-2 B f_1
			-\frac{B^2}{(a+n \hbar )^2}-a \lambda  \hbar -\lambda  \left[ (a+n \hbar )^2-a^2 \right] +\frac{\lambda  \hbar ^2}{4}
	}}{f_1^2-\lambda}
	~~.
\end{equation}
The integration limits are the classical turning points for a particle with energy $E_n$ in the potential $V(x)$ from Eq. (\ref{eq:L}); i.e., they are the roots of 
\begin{equation}\label{eq:TurningPts_classIIB}
	-\left( a-\frac\hbar2\right)^2 f_1^2-2 B f_1
	-\frac{B^2}{(a+n \hbar )^2}-a \lambda  \hbar -\lambda  \left[ (a+n \hbar )^2-a^2 \right] +\frac{\lambda  \hbar ^2}{4} = 0~.
\end{equation}
Since these roots depend on $\lambda$ and $a$,  we have the following three cases, as listed in Table. (\ref{table:conventional-II}):
\begin{enumerate}
	\item[] IIB1) $ \lambda <0$ and $a>\hbar/2$, 
	\item[] IIB2) $ \lambda >0$ and $a<0$, 
	\item[] IIB3) $ \lambda >0$ and $a>\hbar/2$.
\end{enumerate}

\medskip
\noindent
\textbf{Case IIB1:} In this case we have $ \lambda <0$ and $a>\hbar/2$. Without loss of generality, we  set $ \lambda=-1$. The roots of Eq. (\ref{eq:TurningPts_classIIB}) are then given by
\begin{equation}\label{eq:TurningPts_classIIB1}
	\left( \begin{array}{c} f_{1L} \\ f_{1R} \end{array}\right) 
	=
	\frac{-4 B (a+ n\hbar) \mp \sqrt{{\hbar (2 n+1) (4 a+(2 n-1)\hbar ) \left((\hbar-2 a)^2 
				(a+ n\hbar)^2+4 B^2\right)}}}{(\hbar-2 a)^2(a+ n\hbar)}
	~.
\end{equation}
Using the positivity of the integrand between the turning points, we write Eq. (\ref{eq:I_classIIB}) as 
\begin{equation}\label{eq:I_classIIB1}
	I(a,n,\hbar)= \left( a - \frac \hbar2 \right)
	\int_{f_{1L}}^{{f_{1R}}} ~
	\sqrt{ \left(    f_{1R} - f_1 \right) \left(    f_1 - f_{1L}  \right)  }
	\quad 
	\frac{df_1}{1+f_1^2}~~,
\end{equation}
where we have set ${f_1^\prime}=f_1^2+1$. This integral is of the form $I_3(y_1,y_2)$ in the Appendix, hence: 
\begin{equation}\label{eq:I_classIIB1a}
	I(a,n,\hbar)= \left( a - \frac h2 \right) \left[ 
	\frac{\pi}{\sqrt{2}}\left[ \sqrt{1+{f_{1L}}^2}  \sqrt{1+{f_{1R}}^2}  -{{f_{1L}} \, {f_{1R}} }+1                  \right]^{1/2} - \pi\right] ~.~
\end{equation}
Substituting $f_{1L}$ and $f_{1R}$ from Eq. (\ref{eq:TurningPts_classIIB1}), we get
\begin{equation} 
	I(a,n,\hbar)= \left( n+ \frac{1}{2} \right) \hbar \pi~.
\end{equation}

\medskip
\noindent
\textbf{Case IIB2:} In this case we have $ \lambda >0$ and $a<0$. From unbroken supersymmetry, $B<a^2$ and $a+n\hbar<0$ for all bound states. This implies that these potentials must have a finite number of bound states. Without loss of generality, we set $ \lambda=1$. The roots of Eq. (\ref{eq:TurningPts_classIIB}) are then given by
\begin{equation}\label{eq:TurningPts_classIIB2}
	\left( \begin{array}{c} f_{1L} \\ f_{1R} \end{array}\right) 
	=\frac{-B \mp \sqrt{B^2+\left( a - \frac \hbar2 \right)^2 \left(-\frac{B^2}{(a+n \hbar )^2}-2 a n \hbar -a \hbar -n^2 \hbar ^2+\frac{\hbar ^2}{4}\right)}}{\left( a - \frac \hbar2 \right)^2}
	~.
\end{equation}
Using the positivity of the integrand between the turning points, Eq. (\ref{eq:I_classIIB}) becomes 
\begin{equation}\label{eq:I_classIIBx}
	I(a,n,\hbar)= \left(-a + \frac\hbar2 \right) 
	\int_{f_{1L}}^{{f_{1R}}} ~
	\sqrt{\left(    f_{1R} - f_1 \right) \left(    f_1 - f_{1L}  \right)   }
	~
	\frac{df_1}{1-f_1^2}~~,
\end{equation}
where $ 1-f_1^2 $ comes from substituting $f_1^\prime = f_1^2-1$. Note that $f_1^2<\lambda = 1$, and $\left(-a + \frac\hbar2 \right)>0$. Thus the WKB integral is positive. From the Appendix, this integral yields
\begin{equation} 
	I(a,n,\hbar) =
	\frac{1}{2} \pi \left(-a + \frac\hbar2 \right) \left[2-\sqrt{({f_{1L}}-1) ({f_{1R}}-1)}-\sqrt{({f_{1L}}+1) ({f_{1R}}+1)}\right]
	~,
\end{equation}
which from Eq. (\ref{eq:TurningPts_classIIB2}) gives
\begin{equation} 
	I(a,n,\hbar)= \left( n+ \frac{1}{2} \right) \hbar \pi~.
\end{equation}

\medskip
\noindent
\textbf{Case IIB3:} In this case we have $ \lambda >0, a>\hbar/2$, and  $f_1^2>\lambda$. Without loss of generality, we set  $\lambda=1$ and choose $f_1<-1$. Then, the roots of Eq. (\ref{eq:TurningPts_classIIB}) are
\begin{equation}\label{eq:TurningPts_classIIB3}
	\left( \begin{array}{c} f_{1L} \\ f_{1R} \end{array}\right) 
	=\frac{-B \mp \sqrt{B^2+ \left( a - \frac \hbar2 \right)^2 \left(-\frac{B^2}{(a+n \hbar )^2}-2 a n \hbar -a \hbar -n^2 \hbar ^2+\frac{\hbar ^2}{4}\right)}}{\left( a - \frac \hbar2 \right)^2}
	~,
\end{equation}
and the WKB integral of Eq. (\ref{eq:I_classIIB}) can be written as 
\begin{equation}\label{eq:I_classIIBx}
	I(a,n,\hbar)= \left(a - \frac\hbar2 \right) 
	\int_{f_{1L}}^{{f_{1R}}} ~
	\sqrt{ \left(    f_1 - f_{1L}  \right) \left(    f_{1R} - f_1 \right)  }
	\quad 
	\frac{df_1}{f_1^2-1}~~.
\end{equation}
From the Appendix, this integral is 
\[I(a,n,\hbar)= \left(a - \frac\hbar2 \right) \frac{1}{2} \pi  \left(\sqrt{({f_{1L}}-1) ({f_{1R}}-1)}-\sqrt{({f_{1L}}+1) ({f_{1R}}+1)}-2\right)~.
\]
Substituting for the roots from Eq. (\ref{eq:TurningPts_classIIB3}) yields
\begin{equation} 
	I(a,n,\hbar)= \left( n+ \frac{1}{2} \right) \hbar \pi~.
\end{equation}

\subsection{Class III}\label{subsec:classIII}

As we saw in Sec. \ref{Sec:CompleteList}, the superpotential for this class is of the form $W=a f_1+f_2$. In this case, $f_1^\prime=f_1^2 -\lambda$ and $f_2^\prime = f_1 f_2- \varepsilon$, where $\lambda$ and $\varepsilon$ are constants. We now consider separately subclasses IIIA, where $\lambda=0$, and IIIB, where $\lambda\neq0$.

\subsubsection{Subclass IIIA:  $\lambda=0$}

Since $\lambda=0$,   $f_1^\prime=f_1^2$ and $f_1$ cannot be zero for any $x$.  Without loss of generality, we choose $f_1<0$. For unbroken supersymmetry, we require $a>0$ \cite{Gangopadhyaya2020}. Also,
\begin{equation}
	\label{eq:3a}
	f_2^\prime = f_1 f_2- \varepsilon~.
\end{equation}
The homogeneous part of Eq. (\ref{eq:3a}) is solved by $f_2=\alpha f_1$, where $\alpha$ is a constant. A particular solution is $f_2=\epsilon/\left(2 f_1\right)$. Therefore, with a redefinition of parameters $a+\alpha \to a$ and $\varepsilon \to -\omega$, we rewrite $W=a f_1 -\omega /\left(2 f_1\right)$.

From Eq. \ref{PDE1}, we have
\begin{equation}
	\label{eq:g3a}
	\frac{dg}{da} = 2\omega~.
\end{equation}
To avoid level-crossing, $dg/da > 0$, so  $\omega > 0$. The energy eigenvalues are 
\[
E_n=2 n \hbar \omega
~.
\]
Therefore Eq. (\ref{eq:wkb0}) becomes
\begin{equation}\label{eq:I_classIIIA}
	I(a,n,\hbar) =
	\int_{f_{1L}}^{{f_{1R}}}
	\sqrt{2n \hbar \omega -
		\left(a f_1-\frac{\omega}{2 f_1}\right)^2 +\hbar\left( a f_1^2 + \frac{\omega}{2}\right)  - \frac14 \hbar^2 f_1^\prime}\quad 
	\frac{df_1}{f_1^2}~~,
	\end{equation}
where the limits $f_{1L}$ and $f_{1R}$ are the roots of $2n \hbar \omega -
\left(a f_1-\frac{\omega}{2 f_1}\right)^2 +\hbar\left( a f_1^2 + \frac{\omega}{2}\right)  - \frac14 \hbar^2 f_1^\prime$.
Using the fact that $f_1^\prime=f_1^2$, this becomes 
\begin{equation}
	I(a,n,\hbar) =
	-\int_{f_{1L}}^{{f_{1R}}}
	\sqrt{2n \hbar \omega f_1^2 -
		\left(a f_1^2 -\frac{\omega}{2}\right)^2 +\hbar\left( a f_1^4 + \frac{\omega f_1^2}{2}\right)  - \frac14 \hbar^2 f_1^4}\quad 
	\frac{df_1}{f_1^3}~~.
\end{equation}

We now change integration variable such that $y=f_1^2$. Since $f_1<0$, the right and left turning points for the integral are given by $y_L=f_{1R}^2$ and $y_R=f_{1L}^2$, respectively. This yields
\begin{equation}
	\label{eq:57}
	I(a,n,\hbar) =
	\int_{y_{L}}^{{y_{R}}}
	\sqrt{-\left(a-\hbar/2\right)^2 y^2  + \left(2n \hbar \omega + a \omega  + \hbar \omega /2\right) y -\omega^2/4\quad} 
	\frac{dy}{2y^2}~~,
\end{equation}
where the turning points are given by
\begin{equation}\label{eq:TurningPts_classIIIA}
	\left( \begin{array}{c} y_{L} \\ y_{R} \end{array}\right) 
	=
	\frac{2 a \omega + \hbar \omega + 4 n \hbar \omega \mp 2\sqrt{2}\sqrt{a \hbar \omega^2 + 2 a n \hbar \omega^2 + \hbar^2 n \omega^2 + \hbar^2 n \omega^2 + 2 \hbar^2 n^2 \omega^2}}{(2a-\hbar)^2}
	~.
\end{equation}
As shown in Eq. (\ref{eq:a>h}), this case requires $a > \hbar/2$ for the proper behavior of the wave function near the singular points. Using the fact that $y=f_{1}^2>0$, the integral of Eq. (\ref{eq:57}) can be written in the form of $I_{2b}$ in  the Appendix: 
\begin{equation}\label{eq:I_classIIIA}
	I(a,n,\hbar)=\frac{1}{2}\left( a - \frac {\hbar}{2} \right)	\int_{y_{L}}^{{y_{R}}} 
	\frac{\sqrt{ \left(    y_{R} - y \right) \left(    y - y_{L}  \right)   }}{y^2}
	\quad dy~~,
\end{equation}
which integrates to
\begin{equation}
	I(a,n,\hbar)=\frac{\pi}{4}\left(a-\frac{\hbar}{2}\right)\frac{y_L+y_R-2\sqrt{y_L y_R}}{2\sqrt{y_L y_r}}.
\end{equation}
Substituting the turning points yields
\begin{equation}
	I(a,n,\hbar)=\left(n+\frac 12\right)\hbar\pi.
\end{equation} 

\subsubsection{Class IIIB:  $\lambda\neq0$}
For Class IIIB, $f_1^\prime=f_1^2 - \lambda$, $f_2^\prime = f_1 f_2-\epsilon$, and $W = af_1 + f_2$. As previously noted \cite{Gangopadhyaya2020, Gangopadhyaya2021}, the homogeneous and particular solutions for $f_2$ are given by $B \sqrt{\left|f_1^2-\lambda\right|}$ and $\left(\epsilon/\lambda\right)f_1$, respectively. Redefining the parameter $a$, we get $W=a f_1 + B\sqrt{\left|f_1^2-\lambda\right|}$. The eigenenergies are $E_n=\lambda\left[a^2-\left(a+n\hbar\right)^2\right]$, where $\lambda\left( a + n\hbar \right) < 0$ to avoid level crossing.

Note that  if $f_1^2$ were to equal $\lambda$ anywhere in the domain, all derivatives of $f_1$ would be zero at that point and $W$ would be constant. Therefore, $f_1^2-\lambda$ cannot change sign within the domain. We now consider three cases, as listed in Table. \ref{table:conventional-III}: 
\begin{enumerate}
	\item[] IIIB1) $ \lambda <0$, 
	\item[] IIIB2) $ \lambda >0$ and $f_1^2<\lambda$, 
	\item[] IIIB3) $ \lambda >0$ and $f_1^2>\lambda$.
\end{enumerate}

\medskip
\noindent
\textbf{Case IIIB1:} In this case, $\lambda <0$. 
For simplicity, we define $\alpha=-\lambda>0$.	
To simplify the calculation we introduce the functions ${\cal S}$ and ${\cal C}$ as done previously in \cite{Ballesteros1993}. We define  $y\equiv\frac{\sqrt{\lambda}-f_1}{\sqrt{\lambda} + f_1}=\frac{i\sqrt{\alpha}-f_1}{i\sqrt{\alpha} + f_1}$,  ${\cal S}\equiv\frac{y^{1/2}-y^{-1/2}}{2\sqrt{\lambda}}=\frac{y^{1/2}-y^{-1/2}}{2i\sqrt{\alpha}}$, and ${\cal C}\equiv\frac{y^{1/2}+y^{-1/2}}{2}$. In this case, $y$ is complex, while  ${\cal C}$ and ${\cal S}$ are real. With these definitions, the functions satisfy the following identities:
$$
\begin{array}{lll}
	{d{\cal{C}}}/{dx}=-\alpha {\cal{S}}~, & {d{\cal{S}}}/{dx}={\cal{C}}~, &
	{\cal{C}}^2(y)+\alpha \,{\cal{S}}^2(y) = 1~.
\end{array}
$$
The energy eigenvalues are given by 
\[
E_n= n \hbar \left(2a + n \hbar\right)\alpha
~,
\]
where $a>0$ to avoid level crossing.

With these definitions and identities, we  write 
\begin{equation}
	W(x)=a\frac{\alpha {\cal S}(x)}{{\cal C}(x)} + \frac{B}{{\cal C}(x)}~,
\end{equation}
where $B$ is a constant. Equation (\ref{eq:wkb0}) becomes 
\begin{eqnarray}\label{eq:I_classIIIB1a}
	I(a,n,\hbar) & = &
	\int_{x_L}^{{x_R}}
	\left[n \hbar \alpha \left(2a+n \hbar\right)-\left(\frac{B}{{\cal C}(x)}+\frac{a {\cal S}(x)\alpha}{{\cal C}(x)}\right)^2  \right.  \nonumber\\
	&+&	\left. \hbar\left(a \alpha + \frac {B{\cal S}(x)
		\alpha}{{\cal C}^2} + \frac{a {\cal S}^2(x) \alpha^2}{{\cal C}^2}\right)- \frac{\hbar^2}{4}\left( \frac{{\cal S}^2(x)\alpha^2}{{\cal C}^2(x)}+  \alpha \right)\right]^{1/2}dx.
\end{eqnarray}
Without loss of generality, we take $\alpha=1$. Using the facts that ${\cal C}^2=1-{\cal S}^2$ and $d{\cal S}/{dx}={\cal C}~$, we change the integration variable to $S$:
\begin{eqnarray}\label{eq:IIIB1Sint}
	I(a,n,\hbar) & = &
	\int_{{\cal S}_{L}}^{{{\cal S}_{R}}}
	\frac{d{\cal S}}{2(1-{\cal S}^2)}\times \nonumber\\
	&&
	\sqrt{-4\left(a+n\hbar\right)^2{\cal S}^2
		-4B(2a-\hbar){\cal S} -4B^2 +\hbar(1+2n)(4a + \hbar(2n-1)}.	
\end{eqnarray}
The integration limits are given by
\begin{equation}\label{eq:TurningPts_classIIIB1}
	\left( \begin{array}{c} {\cal S}_{L} \\ {\cal S}_{R} \end{array}\right) 
	=
	\frac{-2 a B +B\hbar \mp\sqrt{\hbar(1+2n)(a-B-n\hbar)(a+B+n\hbar)(4a-\hbar+2 n \hbar)}}{2(a+n\hbar)^2}~.
\end{equation}
Equation (\ref{eq:IIIB1Sint}) becomes 
\begin{equation}\label{eq:I_classIIIB1}
	I(a,n,\hbar)=\left( a + n\hbar \right)	\int_{{\cal S}_{L}}^{{{\cal S}_{R}}} 
	\frac{\sqrt{ \left( {\cal S}_{R} - {\cal S}\right) \left(  {\cal S} - {\cal S}_{L}  \right) }}{1-{\cal S}^2}
	~ d{\cal S}~.
\end{equation}

As previously stated, ${\cal S}$ and ${\cal C}$ are both real and ${\cal S}^2 + {\cal C}^2$ = 1; note also that $f_1=\alpha {\cal S}/{\cal C}$, so for finite $f_1$, we must have ${\cal C}\neq 0$. Therefore, $-1 < {\cal S} < 1$.  The integral in Eq. (\Ref{eq:I_classIIIB1}) is therefore of the form $I_4$ in the Appendix. The solution is
\begin{equation}
		I(a,n,\hbar)
		 =\frac{\pi}{2}\left[2- \sqrt {\left( 1-{\cal S}_L\right) \left( 1-{\cal S}_R\right)}
		- \sqrt {\left( 1+{\cal S}_L\right) \left( 1+{\cal S}_R\right)}
		\right]~.
		\end{equation}
Recall that proper behavior of the wavefunction requires $a>|B|+\hbar/2$ for this case. 
With this requirement, substituting the integration limits yields 
\begin{equation}
	I(a,n,\hbar)=\left(n+\frac 12\right)\hbar\pi~.
\end{equation} 

\medskip
\noindent
\textbf{Case  IIIB2:} In this case we have $\lambda>0$, $f_1^2<\lambda$. We again use $y\equiv\frac{\sqrt{\lambda}-f_1}{\sqrt{\lambda} + f_1}$,  ${\cal S}\equiv\frac{y^{1/2}-y^{-1/2}}{2\sqrt{\lambda}}$, and ${\cal C}\equiv\frac{y^{1/2}+y^{-1/2}}{2}$. Since $f_1^2 <\lambda$, we have $-\sqrt{\lambda}<f_1<\sqrt{\lambda}$, $y>0$, ${\cal C}>0$, and ${\cal S}\in\mathbb{R}$. Therefore:
	$$
	\begin{array}{lll}
		{d{\cal{C}}}/{dx}=\lambda {\cal{S}}~, & {d{\cal{S}}}/{dx}={\cal{C}}~, &
		{\cal{C}}^2-\lambda \,{\cal{S}}^2 = 1~.
	\end{array}
	$$
The energy eigenvalues are given by 
\[
E_n=- n \hbar \left(2a + n \hbar\right)\lambda
~.
\]
To avoid level-crossing, $a+n\hbar<0$ for all bound states. Consequently, we  write 
	\begin{equation}
		W(x)=-a\frac{\lambda {\cal S}(x)}{{\cal C}(x)} + \frac{B}{{\cal C}(x)} ~,
	\end{equation}
where $B$ is constant. Equation (\ref{eq:wkb0}) then becomes 
	\begin{eqnarray}\label{eq:I_classIIIB2}
		I(a,n,\hbar) & = &
		\int_{x_L}^{x_R}
		\left[-n \hbar \lambda \left(2a+n \hbar\right)-\left(\frac{B}{{\cal C}(x)}-\frac{a {\cal S}(x)\lambda}{{\cal C}(x)}\right)^2  \right. + \nonumber\\
		&&	\left. \hbar\left(-a \lambda - \frac {B{\cal S}(x)
			\lambda}{{\cal C}^2} + \frac{a {\cal S}^2(x) \lambda^2}{{\cal C}^2}\right)- \frac{\hbar^2}{4}\left( \frac{{\cal S}^2(x)\lambda^2}{{\cal C}^2(x)}-  \lambda \right)\right]^{1/2}dx.
	\end{eqnarray}
Because $\lambda>0$, we set $\lambda=1$. Changing the integration variable to ${\cal S}$, we obtain
\begin{eqnarray}\label{eq:I_classIIIB1_S}
		I(a,n,\hbar) & =& \int_{{\cal S}_L}^{{\cal S}_R} \frac{d{\cal S}}{2\left(1+  {\cal S}^2\right)} 
		\times 	
		\nonumber\\
		&&
		 \left\{
		 -4B^2 + \hbar\left[\hbar-4\left(a+2a n + \hbar n^2\right)\right] - 4B\left(-2a +\hbar\right){\cal S} -4\left(a+n \hbar\right)^2{\cal S}^2 \right\}^{1/2}\nonumber,
	\end{eqnarray}
where the integration limits are given by 
\begin{equation}\label{eq:TurningPts_classIIIB2}
	\left( \begin{array}{c} {\cal S}_{L} \\ {\cal S}_{R} \end{array}\right) 
	=
	\frac{2 a B- B\hbar\mp \sqrt{-\hbar(1+2n)(4a-\hbar + 2 n \hbar)\left[B^2+(a+n \hbar)^2\right]}}{2(a+ n \hbar)^2}~.
	\end{equation}
Since we are integrating between zeros of the numerator, the integrand can be factored to yield
\begin{eqnarray}\label{eq:I_classIIIB3_C}
	I(a,n,\hbar) =-\left(a+n \hbar\right) \int_{{\cal S}_{L}}^{{\cal S}_R}\frac{\sqrt{
		\left({\cal S}-{\cal S}_L\right)\left({\cal S}_R-{\cal S}\right)}}	{{\cal S}^2 +1}\nonumber~.
\end{eqnarray}
This is of the form of the integral $I_3$ from the Appendix, and the solution is
\begin{equation}
I(a,n,\hbar) =	-\left(a+n \hbar\right)\left[\frac{\pi}{\sqrt{2}}\left(\sqrt{1+{\cal S}_L^2}\sqrt{1+{\cal S}_R^2}-{\cal S}_L {\cal S}_R+1\right)^{1/2}-\pi\right]~,
\end{equation}
which yields
\begin{equation} 
	I(a,n,\hbar)= \left( n+ \frac{1}{2} \right) \hbar \pi~.
\end{equation}

\medskip
\noindent
\textbf{Case IIIB3:} We have $\lambda>0$, $f_1^2>\lambda$. We define  $y\equiv\frac{f_1-\sqrt{\lambda}}{f_1+\sqrt{\lambda}}$,  ${\cal S}\equiv\frac{y^{1/2}-y^{-1/2}}{2\sqrt{\lambda}}$, and ${\cal C}\equiv\frac{y^{1/2}+y^{-1/2}}{2}$. Since $f_1^2>\lambda$, either $f_1>\sqrt{\lambda}$ or $f_1<-\sqrt{\lambda}$.  Without loss of generality, we choose  $f_1<-\sqrt{\lambda}.$  Thus $y>1$, ${\cal C}>1$, and ${\cal S}>0$. With these definitions, the functions satisfy the following identities:
	$$
	\begin{array}{llll}
		{d{\cal{C}}}/{dx}=\lambda {\cal{S}}~, & {d{\cal{S}}}/{dx}={\cal{C}}~, &
		{\cal{C}}^2-\lambda \,{\cal{S}}^2 = 1~.
	\end{array}
	$$
In this case, we can write $W=\left(-a\, {\cal C}/{\cal S}+B/{\cal S}\right)$ for some constant $B$. 
From Eq. \ref{PDE1}, we have
\begin{equation}
	\label{eq:g3a}
	\frac{dg}{da} = - 2 a \lambda,
\end{equation}
and hence the energy eigenvalues are given by 
\[
E_n=- n \hbar \left(2a + n \hbar\right)\lambda
~.
\]
To avoid level-crossing, $a+n \hbar < 0$, which implies that these potentials must have a finite number of bound states.
Since ${\cal C}>1$ and ${\cal S}>0$, unbroken supersymmetry requires $  B <a+n \hbar <0$.
In this case, 
Eq. (\ref{eq:wkb0}) becomes
	\begin{eqnarray}\label{eq:I_classIIIB3}
	I(a,n,\hbar) & = &
	\int_{f_{1L}}^{{f_{1R}}}
	\left[-n \hbar \lambda \left(2a+n \hbar\right)-\left(\frac{B}{{\cal S}(x)}-\frac{a\, {\cal C}(x)}{{\cal S}(x)}\right)^2  \right. + \nonumber\\
	&&	\left. \hbar\left(-a \lambda - \frac {B{\cal C}(x)}{{\cal S}(x)^2} + \frac{a\, {\cal C}^2(x) }{{\cal S}^2 (x)}\right)- \frac{\hbar^2}{4}\left( \frac{{\cal C}^2(x)}{{\cal S}^2(x)}-\lambda\right)\right]^{1/2}df_1~.
\end{eqnarray}
Since $\lambda>0$, we set $\lambda=1$. Changing the integration variable to ${\cal C}$, we obtain
	\begin{eqnarray}\label{eq:I_classIIIB3_C}
	I(a,n,\hbar) & = & \nonumber\\
	&&\int_{{\cal C}_{L}}^{{\cal C}_R}
	\frac {\sqrt{-4B^2  + 4 a \hbar -\hbar^2 + 8 a n \hbar + 4 \hbar^2 n^2 + 4 B {\cal C}\left(2 a - \hbar \right) - 4 {\cal C}^2\left(a+n \hbar\right)^2}}	{2\left({\cal C}^2 -1\right)}~d{\cal C}\nonumber,
\end{eqnarray}
where the integration limits are
\begin{equation}\label{eq:TurningPts_classIIIB2}
	\left( \begin{array}{c} {\cal C}_{L} \\ {\cal C}_{R} \end{array}\right) 
=\frac{2 a B - B \hbar \mp \sqrt{\hbar(1+ 2n)(a-B+n\hbar)(a+B+n\hbar)(4a-\hbar+2n\hbar)}}{2(a+n\hbar)^2}~.
\end{equation}
The integrand can be factored to yield
\begin{eqnarray}\label{eq:I_classIIIB3_C}
	I(a,n,\hbar) =-\left(a+n \hbar\right)\int_{{\cal C}_{L}}^{{\cal C}_R}\frac{\sqrt{
	\left({\cal C}-{\cal C}_L\right)\left({\cal C}_R-{\cal C}\right)}}	{{\cal C}^2 -1}\nonumber~.
\end{eqnarray}
Since ${\cal C}>0$, this is of the form of integral $I_{5a}$ from the Appendix and its solution is given by
\begin{equation}
	I(a,n,\hbar) =-\left(a+n \hbar\right)\frac{\pi}{2}\left(\sqrt{\left({\cal C}_L +1\right)\left({\cal C}_R+1\right)}-\sqrt{\left({\cal C}_L -1\right)\left({\cal C}_R-1\right)}-2\right).
\end{equation}
Substituting the values of the turning points and requiring $a - B -\hbar/2 >0$, this yields
\begin{equation}
	I(a,n,\hbar)=\left(n+\frac 12\right)\hbar\pi.
\end{equation}

\section{Interrelation between SWKB and Langer-corrected WKB}\label{sec:swkb}

In Sec.~\ref{glc}, we proved that shape invariance leads to exactness of WKB with the generalized Langer correction for all conventional potentials. In Ref.~\cite{Gangopadhyaya2020}, the authors proved that shape invariance also ensures the exactness of the SWKB quantization condition for the same potentials. In this section we show that the requirements for shape-invariance of conventional potentials given by Eqs.~(\ref{PDE1}) and (\ref{PDE2}) interconnect the two formalisms: SWKB and Langer-corrected WKB. 

We begin by integrating Eq. (\ref{PDE1}):
\begin{equation}
	\label{eq:VL-SWKB1}
	\int_{a-\hbar/2}^a \frac{\partial}{\partial a} \Big( W^2(x,a) + g(a) \Big) da = 2 \int_{a-\hbar/2}^a \frac{\partial }{\partial x} W(x,a)\, da~.
\end{equation}
Then, from Eq. (\ref{Wform}) we obtain
\begin{equation}
	\label{eq:VL-SWKB2}
	W^2(x,a) - \hbar \Big(af_1'(x) +f_2'(x) \Big) + \frac 14 \hbar^2 f_1'(x) + g(a)
	= W^2(x,a-\hbar/2) + g(a-\hbar/2)~.
\end{equation}
Subtracting $g(a+n\hbar)$ from both sides of Eq. (\ref{eq:VL-SWKB2}) and using Eq. (\ref{eq:En}) we get
\begin{equation}
	\label{eq:VL-SWKB3}
	E_n(a) - \left( V_-(x,a) +  \frac 14 \hbar^2 f_1'(x) \right) 
	= E_{n+ \frac 12}(\tilde{a}) - W^2(x,\tilde{a}) ~,
\end{equation}
where $\tilde{a} \equiv a-\hbar/2$. Note that $V_-(x,a) +  \frac 14 \hbar^2 f_1'(x)$ is exactly the  generalized Langer-corrected potential we conjectured in Eq. (\ref{eq:gen-Langer}). We have proven in this paper that the integral of the square root of the LHS of Eq. (\ref{eq:VL-SWKB3}) is $(n+\frac 12)\hbar\pi$.  The square root of the RHS of the same equation is the integrand of the SWKB condition Eq. (\ref{eq:swkb}) with $n\to n+ \frac 12$. Consequently, these two formalisms are interrelated.  

\section{Conclusion}\label{conc}

Conventional potentials \cite{Bougie2010} play an important role in SUSYQM due to their shape invariance. Being exactly solvable, they provide a good testing ground for approximation methods, especially for non-perturbative approximations such as the semiclassical WKB quantization. 

In this paper we introduced a generalized Langer correction of the form $\frac{\hbar^2 f_1’}{4}$ and proved that the semiclassical WKB quantization condition yields exact eigenenergies for all conventional potentials. Note that this correction is needed even for some potentials which are defined over the entire real axis, such as Class IIB2 and Class IIIB2 potentials. Therefore, as also suggested by previous authors \cite{Froman2004},  the rationale of Langer \cite{Langer} regarding the transformation of the domain from semi-infinite to infinite, does not appear to play a fundamental role in ensuring WKB-exactness.

The conventional shape-invariant potentials share several remarkable properties, all of which arise from the form of these potentials. First, they are all solvable, yielding analytic solutions for their eigenenergies \cite{Dutt_SUSY,Cooper-Khare-Sukhatme,Gangopadhyaya-Mallow-Rasinariu}. Second, they are all SWKB exact without the need for any correction in both the broken \cite{Gangopadhyaya2021} and unbroken \cite{Gangopadhyaya2020} phases. Finally, as we have shown in this manuscript, they all become WKB exact with the addition of a generalized Langer correction of the form $\frac{\hbar^2 f_1’}{4}$.  In this paper, we have demonstrated that the SWKB-exactness and the WKB-exactness are interrelated. The relationships between these properties are interesting and merit further study.

\begin{acknowledgments}
	The authors would like to thank Arturo Ramos for in-depth and fruitful discussions. We also gratefully acknowledge the anonymous referees for their thoughtful and helpful comments. 
\end{acknowledgments}

\newpage

\appendix{Appendix: List of Relevant Integrals}

\noindent
Note that many of these integrals previously appeared in \cite{Hruska1997}.
\begin{eqnarray}
	I_0(y_1,y_2) &\equiv& \int_{y_1}^{y_2} ~dy~
	\sqrt{\left( y_2-y\right)\left( y-y_1\right)}
	=\frac{\pi}{8}\left( y_2-y_1\right)^2~;~
	\nonumber \\	
	I_{1a}(y_1,y_2) &\equiv& \int_{y_1}^{y_2} \frac{~dy~
		\sqrt{\left( y_2-y\right)\left( y-y_1\right)}}{y}
	=\frac{\pi}{2}\left( y_1+y_2\right) - \pi\sqrt{y_1 \, y_2}~,~(0<y_1<y_2)~;~
	\nonumber \\	
	I_{1b}(y_1,y_2) &\equiv& \int_{y_1}^{y_2} \frac{~dy~
		\sqrt{\left( y_2-y\right)\left( y-y_1\right)}}{y}
	=\frac{\pi}{2}\left( y_1+y_2\right) + \pi\sqrt{y_1 \, y_2}~,~(y_1<y_2<0)~;~
	\nonumber \\	
	I_{2a}(y_1,y_2) &\equiv&  \int_{y_1}^{y_2}
	\frac{~dy~
		\sqrt{\left( y_2-y\right)\left( y-y_1\right)}}{y^2} =
	-\frac{\pi \left(
		{y_2} \sqrt{{y_1} {y_2}}+{y_1} \left(\sqrt{{y_1} {y_2}}+2 {y_2} \right)
		\right)}
	{2 {y_1} {y_2} }
	~,~(y_1<y_2<0)~;~
	\nonumber \\	
	I_{2b}(y_1,y_2) &\equiv&  \int_{y_1}^{y_2}
	\frac{~dy~
		\sqrt{\left( y_2-y\right)\left( y-y_1\right)}}{y^2} =
	\frac{\pi \left( {y_1}+{y_2} - 2 \sqrt{{y_1} {y_2}}
		\right)}
	{2 \sqrt{{y_1} {y_2}} }
	~,~(0<y_1<y_2)~;~
	\nonumber \\	
	I_3(y_1,y_2) &\equiv& \int_{y_1}^{y_2} \frac{~dy~
		\sqrt{\left( y_2-y\right)\left( y-y_1\right)}}{1+y^2}
	=\frac{\pi}{\sqrt{2}}\left[ \sqrt{1+y_1^2}  \sqrt{1+y_2^2}  -{y_1 \, y_2}+1                  \right]^{1/2} - \pi~;~
	\nonumber \\	
	I_4(y_1,y_2) &\equiv& \int_{y_1}^{y_2} \frac{~dy~
		\sqrt{\left( y_2-y\right)\left( y-y_1\right)}}{1-y^2} =\frac{\pi}{2}\left[2- \sqrt {\left( 1-y_1\right) \left( 1-y_2\right)}
	- \sqrt {\left( 1+y_1\right) \left( 1+y_2\right)}
	\right] ~, \nonumber \\
	&&\qquad \qquad {\rm where}~	(-1<y_1<y_2<1)~;~
	\nonumber \\
	I_{5a}(y_1,y_2) &\equiv& \int_{y_1}^{y_2} \frac{~dy~
		\sqrt{\left( y_2-y\right)\left( y-y_1\right)}}{y^2-1} =\frac{\pi}{2}\left[\sqrt {\left( y_1+1\right) \left( y_2+1\right)}- \sqrt {\left( y_1-1\right) \left( y_2-1\right)} -2
	\right] ~, \nonumber \\
	&&\qquad \qquad{\rm where}~	(1<y_1<y_2)~;~ \nonumber \\
	I_{5b}(y_1,y_2) &\equiv& \int_{y_1}^{y_2} \frac{~dy~
		\sqrt{\left( y_2-y\right)\left( y-y_1\right)}}{y^2-1} =
	\frac{1}{2} \pi  \left(\sqrt{(y_1-1) (y_2-1)}-\sqrt{({y_1}+1) (y_2+1)}-2\right)
	~, \nonumber \\
	&&\qquad \qquad{\rm where}~	(y_1<y_2<-1)~.
	\nonumber \\	
\end{eqnarray}


\begin{thebibliography}{10}
\bibitem{Froman2004} N. Froman, P.O. Froman, {Physical Problems Solved by Phase-Integral Method}, Cambridge University Press, Cambridge, 2004.
On page 61 of the book, authors state ``In this connection it should be remarked that some authors motivate the Langer transformation by saying that the WKB approximation applies to the interval $(-\infty,\infty)$, but not to the interval $(0,\infty)$. This statement is obviously inadequate, since the extent of the interval is not involved in the derivation of the approximation.”
\end{thebibliography}
\end{document}